\documentstyle[aps,prb,epsf,multicol]{revtex}
\tighten
\begin{document}
\draft
\newcommand{\bn}{{\bf n}}
\newcommand{\bp}{{\bf p}}
\newcommand{\br}{{\bf r}}
\newcommand{\bq}{{\bf q}}
\newcommand{\bj}{{\bf j}}
\newcommand{\bE}{{\bf E}}
\newcommand{\eps}{\varepsilon}
\newcommand{\la}{\langle}
\newcommand{\ra}{\rangle}
\newcommand{\cK}{{\cal K}}
\newcommand{\cD}{{\cal D}}
\newcommand{\hp}{\hat p}
\newcommand{\hq}{\hat q}
\newcommand{\hx}{\hat x}
\newcommand{\hH}{{\hat H}_0}
\newcommand{\mybeginwide}{
    \end{multicols}\widetext
    \vspace*{-0.2truein}\noindent
    \hrulefill\hspace*{3.6truein}
}
\newcommand{\myendwide}{
    \hspace*{3.6truein}\noindent\hrulefill
    \begin{multicols}{2}\narrowtext\noindent
}

\title{ Ground State Energy Fluctuations of a System Coupled to a Bath
 }
\author{K.\ E.\ Nagaev$^{1,2}$ and
M.\ B\"uttiker$^1$
}
\address{
  $^1$D\'epartement de Physique Th\'eorique, Universit\'e de Gen\`eve,
  CH-1211 Gen\`eve 4, Switzerland\\
  $^2$ Institute of Radioengineering and Electronics,
  Russian Academy of Sciences, Mokhovaya ulica 11, 101999 Moscow,
  Russia\\}
\date\today
\maketitle
\bigskip
\begin{abstract}
It is often argued that a small
non-degenerate quantum system coupled to a bath
has a fixed
energy in its ground state since a fluctuation in energy
would require an energy supply from the bath. We consider
a simple model of a harmonic oscillator (the system) coupled
to a linear string
and determine the mean squared energy fluctuations. We also analyze
the two time correlator of the energy  and discuss its behavior
for a finite string.

\end{abstract}

\begin{multicols}{2}
\narrowtext
The coupling of two subsystems is a central problem
in many areas of physics.
The interaction of atoms
with the radiation field has been central in the development
of quantum mechanics \cite{gard}. The division into subsystems often
depends on the questions asked: For instance dephasing
in mesoscopic systems investigates the quantum motion of a single
electron in a conductor viewing all other conduction electrons
as the bath \cite{AAK}. In this work we are interested in the
energy fluctuations in a test system coupled to another system.
Of particular interest is the zero-temperature limit.
It is often argued that in the zero-temperature
limit two subsystems can not exchange energy \cite{alein,gavish}.
The argument is
based on the assumption that both subsystems are in their
separate ground state (which they would assume in the absence
of any coupling). Since both the test system and the bath
are in their ground state neither of the two can supply
an energy to the other \cite{alein,gavish}.
Below we consider a simple system
of an oscillator (the test system) coupled to a linear string
(the bath) and investigate the energy fluctuations of the
test system. This is an exactly solvable model which demonstrates
the existence of energy fluctuations in the zero-temperature
limit. These fluctuations are a consequence of the finite coupling
energy between the test system and the bath.

Often the interaction of two subsystems is
treated in terms of the Einstein coefficients
for the absorption and the spontaneous and stimulated emission
of the test system.
In such an approach rates of transitions of the test system
to change energy form one eigenstate $E_n$ to another eigenstate
$E_m$ are investigated. These are inelastic transition
rates which vanish when both systems are in the ground state.
The resulting probabilities are used in a classical stochastic equation
for the probabilities of occupation of the test system.
Only probabilities enter in the description of the reservoir
system interaction and this approach again yields
no energy fluctuations in the ground state. Such a description
is challenged both in Laser physics \cite{gard} and in the more
recent discussions of macroscopic quantum coherence \cite{cl,weiss}.

The role of zero-point fluctuations in mesoscopic conductors is a
hotly debated issue and we cite here only Ref. \onlinecite{natel}
as an entry to the literature. The point of view taken here has
been applied to investigate the persistent current of a small
mesoscopic loop with a quantum dot capacitively coupled to a
transmission line \cite{cedr}. The persistent current is a measure
of the quantum coherence of the ground state. It was found that
the ground state undergoes a crossover form a state with a well
defined persistent current much larger than its mean square
fluctuations to a state in which the magnitude of the persistent
current is much smaller than its mean square fluctuations as the
coupling to the bath increases \cite{cedr}.

Vacuum fluctuations can have important effects
on the system considered. The Lamb shift is a widely appreciated
example. Another example is the Casimir effect. In both of these
examples the effect of the vacuum can be thought of in terms of a
renormalization of the original parameters
characterizing the system \cite{mill}.
In contrast, the energy fluctuations which we discuss here can not
be obtained from the uncoupled system simply by renormalizing
its parameters: After all, a harmonic oscillator in its ground state
does not fluctuate.

The system we consider is a harmonic oscillator
with mass $m_{0}$ and frequency $\omega_{0}$
described by the energy operator
\begin{equation}
 \hH
 =
 \frac{1}{2m_0}
 \hp^2
 +
 \frac{m_0\omega_0^2}{2}
 \hq^2.
\label{H-1}
\end{equation}
The oscillator is coupled to a harmonic bath, which leads to a
dissipation linear in its velocity\cite{weiss}  with a friction
constant $\eta$. The expectation value of the mean energy of the
test oscillator in the ground state of the system is to the linear
order in the friction constant $\eta$
\begin{equation}
\la \hH
\ra
 = \frac{\hbar \omega_{0}}{2} +
 \frac{\hbar\eta}{2\pi}
 \left[
   \ln
   \left(
     \frac{\omega_c}{\omega_0}
   \right)
   -
   1
 \right]
\label{H-8}
\end{equation}
and the mean squared energy $\la \delta\hH^2 \ra \equiv
\la ( \hH - \la \hH \ra )^{2} \ra $ is
\begin{equation}
 \la
   \delta\hH^2
 \ra
 =
 \frac{\hbar^{2}\omega_0\eta}{2\pi}
 \left[
   \ln
   \left(
     \frac{\omega_c}{\omega_0}
   \right)
   -
   1
 \right].
\label{H-9}
\end{equation}
where $\omega_c$ is a (high-frequency) Debye cut-off of the
spectrum of the bath. Thus both the mean energy and the mean
squared fluctuations in energy increase (for weak coupling)
linearly with the coupling constant $\eta$. Note that the mean
squared fluctuations are proportional to $\hbar^{2}$. The result
for the mean energy is well known\cite{weiss}: In contrast, the
result for the energy fluctuations seems novel. Below we extend
Eq. (\ref{H-9}) to all orders in $\eta$ and investigate also the
two time correlator of the energy (a fourth order correlation).
Eqs. (\ref{H-8}) and (\ref{H-9}) are valid for the infinitely long
string. To emphasize that the overall energy (system plus coupling
energy plus bath) is conserved despite the energy fluctuations in
the test system alone, we will later also consider a string with a
finite number of particles only.

The mean squared energy fluctuations can be determined from the
coordinate correlation function of the oscillator,\cite{weiss}
which is conveniently obtained using the classical response of the
oscillator and the fluctuation - dissipation relation at a finite
temperature $T$
$$
 \la\hq(t) \hq(0)\ra
 =
 \frac{\hbar}{2\pi m_0}
 \int\limits_{-\infty}^{\infty}
 d\omega
 \frac{
   \eta\omega\exp(i\omega t)
 }{
   (
     \omega^2 -\omega_0^2
   )^2
   +
   \eta^2\omega^2
 }
$$
\begin{equation}
 \times
 \left[
   \coth
   \left(
     \frac{\omega}{2T}
   \right)
   +
   1
 \right],
 \label{qq-1}
\end{equation}
and then passing to the limit $T=0$. Introducing $\Omega_{+} =
(\omega^{2}_{0} - \eta^{2} /4)^{1/2}$, for $\omega_0
> \eta/2$ and $\Omega_{-} = (\eta^{2} /4 - \omega^{2}_{0})^{1/2}$
for $\omega_0 < \eta/2$, Eq. (\ref{qq-1}) gives at $t = 0$ the
mean squared displacement \cite{weiss}
\begin{equation}
 \la q^2 \ra
 =
 \frac{\hbar}{
   2 m_0 \Omega_{+}
 }
 \left[
    1
    -
    \frac{2}{\pi}
    \arctan
    \left(
       \frac{
         \eta
       }{2\Omega_{+}
       }
    \right)
 \right],
 \label{q^2-3}
\end{equation}
for $\omega_0 > \eta/2$
and
\begin{equation}
 \la q^2 \ra
 =
 \frac{
   \hbar
 }{
   \pi m_0
   \Omega_{-}
 }
 \ln
 \left(
    \frac{
      \eta
      +
      2\Omega_{-}
    }{
      2\omega_0
    }
 \right),
 \label{q^2-4}
\end{equation}
for $ \omega_0 < \eta/2 $.
The energy fluctuations are determined by the fourth order correlations
of the momentum and coordinate. The momentum is related to the
coordinate via
$
 \hp = m_0 \hat{\dot q}
$
and thus the momentum correlations can be reduced to time-derivatives
of coordinate correlations.
We next make use of the fact that the fluctuations are Gaussian
and thus fourth order correlations can be expressed as sums of
products of second order correlations.
Thus we obtain the energy correlator in the form
$$
 \la
  \delta\hH(t) \delta\hH(0)
 \ra
 \equiv
 \la
   \hH(t) \hH(0)
 \ra
 -
 \la
   \hH
 \ra^2
$$
$$
=
 \frac{1}{2} m_0^2
 \left[
   \frac{\partial^2}{\partial t^2}
   \la
     \hq(t)\hq(0)
   \ra
 \right]^2
 +
 m_0^2 \omega_0^2
 \left[
   \frac{\partial}{\partial t}
   \la
     \hq(t)\hq(0)
   \ra
 \right]^2
$$
 \begin{equation}
 +
 \frac{1}{2} m_0^2 \omega_0^4
 \la
   \hq(t)\hq(0)
 \ra ^2.
\label{H-4}
\end{equation}
For the mean squared fluctuations we need the time derivatives
of the correlator at $t =0$. Evaluating the resulting integrals
yields,
\mybeginwide
\begin{equation}
 \la
   \delta\hH^2
 \ra
 =
 \frac{\hbar^2}{2\pi^2}
 \eta^2
 \ln^2
 \left(
   \frac{\omega_c}{\omega_0}
 \right)
 +
 \frac{\hbar}{\pi}
 m_0\eta
 (
   \omega_0^2 - \eta^2/2
 )
 \ln
 \left(
   \frac{\omega_c}{\omega_0}
 \right)
 \la
   q^2
 \ra
 +
 \frac{1}{8}
 m_0^2
 (
   8\omega_0^4 - 4\omega_0^2\eta^2 + \eta^4
 )
 \la
   q^2
 \ra^2
 -
 \frac{\hbar^2}{4}
 \omega_0^2,
\label{H-10}
\end{equation}
\myendwide
where $\la q^2 \ra$ is given by Eq.~(\ref{q^2-3}).
Expanding Eq.~(\ref{H-10})
up to terms linear in $\eta$ gives Eq.~(\ref{H-9}).
The full result Eq. (\ref{H-10}) is shown in Fig. \ref{FIG.1}
together with the mean energy
\begin{eqnarray}
\la \hH \ra = (\hbar \eta/2\pi)
\ln(\frac{\omega_{c}}{\omega_{0}}) + (m_{0}\Omega_{+}^{2})
<q^{2}> \label{H-11}
\end{eqnarray}
of the test oscillator.

So far we considered the system consisting of the oscillator and
the bath in a thermodynamic equilibrium at a given temperature
$T$, which was then set equal to zero. It is also instructive to
trace the origin of the energy fluctuations by using a microscopic
model of the bath and purely quantum-mechanical considerations
without involving any thermodynamic relations.

It is well known that a harmonic bath can be modeled by a
semi-infinite string of identical particles
\cite{cl,weiss,lamb,rubi,grab,ford} with mass $m_h$ and frequency
$\omega_h$. The electrical analog of this system consists of an
LC-oscillator coupled to a transmission line \cite{yurke}. In the
{ continuous limit} where $m_h \to 0$ and $\omega_h \to \infty$,
the string leads to dissipation that is linear in the velocity of
the oscillator with friction constant $\eta =
(m_{h}/m_{0})\omega_{h}$. In this limit, the interaction with the
string does not shift the resonance frequency of the test
oscillator and results only in a finite damping. The role of the
high-frequency cutoff $\omega_c$ is played now by $\omega_h$.

 To address a system with a well defined ground-state energy, we
consider a test system attached to a chain of a finite length
consisting of $N$ oscillators. The Hamiltonian of a system coupled
to a linear harmonic string is conveniently split into the
following three parts
$$
 \hat{\cal H}
 = \hH' + \hat{H}_i + \hat{H}_b,
 \label{H_tot}
$$
where
\begin{equation}
 \hH'
 =
 \frac{1}{2m_0}
 \hp^2
 +
 \frac{1}{2}
 (
   m_0\omega_0^2 + m_h\omega_h^2
 )
 \hq^2,
\label{H_0'}
\end{equation}
\begin{equation}
 \hat{H}_i
 =
 -
 m_h\omega_h^2
 \hq\hx_1,
\label{H_i}
\end{equation}
$$
 \hat{H}_b
 =
 \frac{1}{2m_h}
 \sum\limits_{\alpha=1}^{N}
 \hp_{\alpha}^2
 +
 \frac{1}{2}
 m_h\omega_h^2\hx_1^2
$$

\noindent
\begin{figure}
 \vspace{3mm}
 \centerline{
  \epsfxsize8cm
  \epsffile{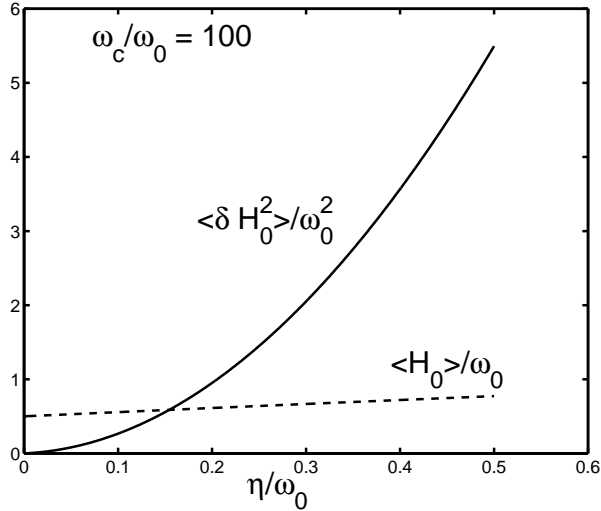}
 }
 \nopagebreak
 \caption{The mean energy $\la \hH \ra$ (broken line)
 and the mean-squared energy fluctuations $\la \delta\hH^2 \ra$
 in the ground state as a function of the coupling constant $\eta$
 in units with $\hbar = 1$.
 }
\label{FIG.1}
\end{figure}
\noindent
\begin{equation}
 +
 \frac{1}{2}
 m_h\omega_h^2
 \sum\limits_{\alpha=1}^{N-1}
 (
   \hx_{\alpha+1} - \hx_{\alpha}
 )^2.
\label{H_b}
\end{equation}
A transition from a finite to the infinite system is most
conveniently traced by watching the behavior of a two-time
correlator of energy. We introduce the exact {\it classical}
normal modes of the total Hamiltonian $\hat{\cal H}$ and calculate
the correlator (\ref{H-4}). Note that this is {\it not} the
quantum- mechanical diagonalization of $\hat{\cal H}$ in the basis
of its eigenstates because the number of these states is
infinitely large even for one oscillator. Instead, we perform a
linear transformation of coordinates $x_{\alpha}$ with an $N+1
\times N+1$ matrix $\hat{U}$ that will simultaneously bring the
quadratic forms
$$
 K
 =
 \frac{1}{2m_0} p^2
 +
 \frac{1}{2m_h}
 \sum\limits_{\alpha}
 p_{\alpha}^2
 =
 \frac{1}{2}
 \sum\limits_{\alpha\beta}
 K_{\alpha\beta}
 p_{\alpha}p_{\beta} ,
$$
$$
 \Pi
 =
 \frac{1}{2}
 \sum\limits_{\alpha\beta}
 \Pi_{\alpha\beta}
 x_{\alpha}x_{\beta}
$$
describing the kinetic and potential energy of the system to a diagonal form.
To this end, we introduce the classical normal modes of the system
$\psi_k$, which obey the set of equations
\begin{equation}
 -\omega_k^2\psi_k(0)
 =
 -
 (
   \omega_0^2 + \mu\omega_h^2
 )
 \psi_k(0)
 +
 \mu\omega_h^2 \psi_k(1),
\label{x_0..}
\end{equation}
\begin{equation}
 -\omega_k^2\psi_k(1)
 =
 \omega_h^2
 [
   \mu \psi_k(0) + \psi_k(2) - 2\psi_k(1)
 ],
\label{x_1..}
\end{equation}
$$
 -\omega_k^2(\alpha)
 =
 \omega_h^2
 [
   \psi_k(\alpha+1) + \psi_k(\alpha-1) - 2\psi_k(\alpha)
 ],
$$
\begin{equation}
 2 \le \alpha < N,
\label{x_alpha..}
\end{equation}
\begin{equation}
 -\omega_k^2\psi_k(N)
 =
 \omega_h^2
 [
   \psi_k(N-1) - \psi_k(N)
 ],
\label{x_N..}
\end{equation}
Here
$\mu = (m_h/m_0)^{1/2}$ is the ratio of the mass of the bath oscillators and
the mass of the test oscillator.
We search the eigenvectors in the form
$$
 \psi_k(0)
 =
 A_k/\mu,
$$
\begin{equation}
 \psi_k(\alpha \ge 1)
 =
 A_k \cos(\lambda_k\alpha)
 +
 B_k \sin(\lambda_k\alpha).
\label{psi_k(alpha>1)}
\end{equation}
On applying the transformation (\ref{psi_k(alpha>1)}) to the Hamiltonian
$\hat{\cal H}_L$, it assumes the form
\begin{equation}
 \hat{\cal H}_L
 =
 \frac{1}{2}
 \sum\limits_{k=1}^{N+1}
 \left(
   \frac{1}{m_h}
   \hat\pi_k^2
   +
   m_h\omega_k^2
   \hat\xi_k^2
 \right).
\label{H-transformed}
\end{equation}
The quantum-mechanical coordinate of the test oscillator $\hat{q}$ can be
written in terms of the normal coordinates $\hat\xi_k$ of the transformed
Hamiltonian $\hat{\cal H}$ (\ref{H-transformed})
\begin{equation}
 \hat{q}(t)
 =
 \sum\limits_{k=1}^{N+1}
 A_k \hat\xi_k(t),
\label{q(t)}
\end{equation}
and the latter can be presented in the form
\begin{equation}
 \hat\xi_k(t)
 =
 \frac{
  \hbar^{1/2}
 }{
  \sqrt{2m_h\omega_h}
 }
 \left(
   \hat{a}_k e^{-i\omega_k t}
   +
   \hat{a}_k^+ e^{i\omega_k t}
 \right),
\label{xi_k(t)}
\end{equation}
where $\hat{a}_k$ and $\hat{a}_k^+$ are time-independent annihilation and
creation operators with the standard commutation rules.

Now one easily obtains from Eq. (\ref{H-4}) an expression for the correlation
function of energy fluctuations
\mybeginwide
$$
 C(t)
 \equiv
 \frac{1}{2}
 \la
   \delta\hH(t) \delta\hH(0)
   +
   \delta\hH(0) \delta\hH(t)
 \ra
 =
 \frac{\hbar^2}{8}
 \frac{m_0^2}{m_h^2}
 \left[
   \left(
     \sum\limits_k
     \omega_k A_k^2
     \cos\omega_k t
   \right)^2
   -
   \left(
     \sum\limits_k
     \omega_k A_k^2
     \sin\omega_k t
   \right)^2
 \right]
$$

\begin{equation}
 +
 \frac{\hbar^2}{4}
 \frac{m_0^2\omega_0^2}{m_h^2}
 \left[
   \left(
     \sum\limits_k
     A_k^2
     \sin\omega_k t
   \right)^2
   -
   \left(
     \sum\limits_k
     A_k^2
     \cos\omega_k t
   \right)^2
 \right]
 +
 \frac{\hbar^2}{4}
 \frac{m_0^2\omega_0^4}{m_h^2}
 \left[
   \left(
     \sum\limits_k
     \frac{ A_k^2 }{\omega_k}
     \cos\omega_k t
   \right)^2
   -
   \left(
     \sum\limits_k
     \frac{ A_k^2 }{\omega_k}
     \sin\omega_k t
   \right)^2
 \right].
\label{ups}
\end{equation}
%
\begin{figure}
 \vspace{3mm}
 \centerline{
\epsfxsize15cm
  \epsffile{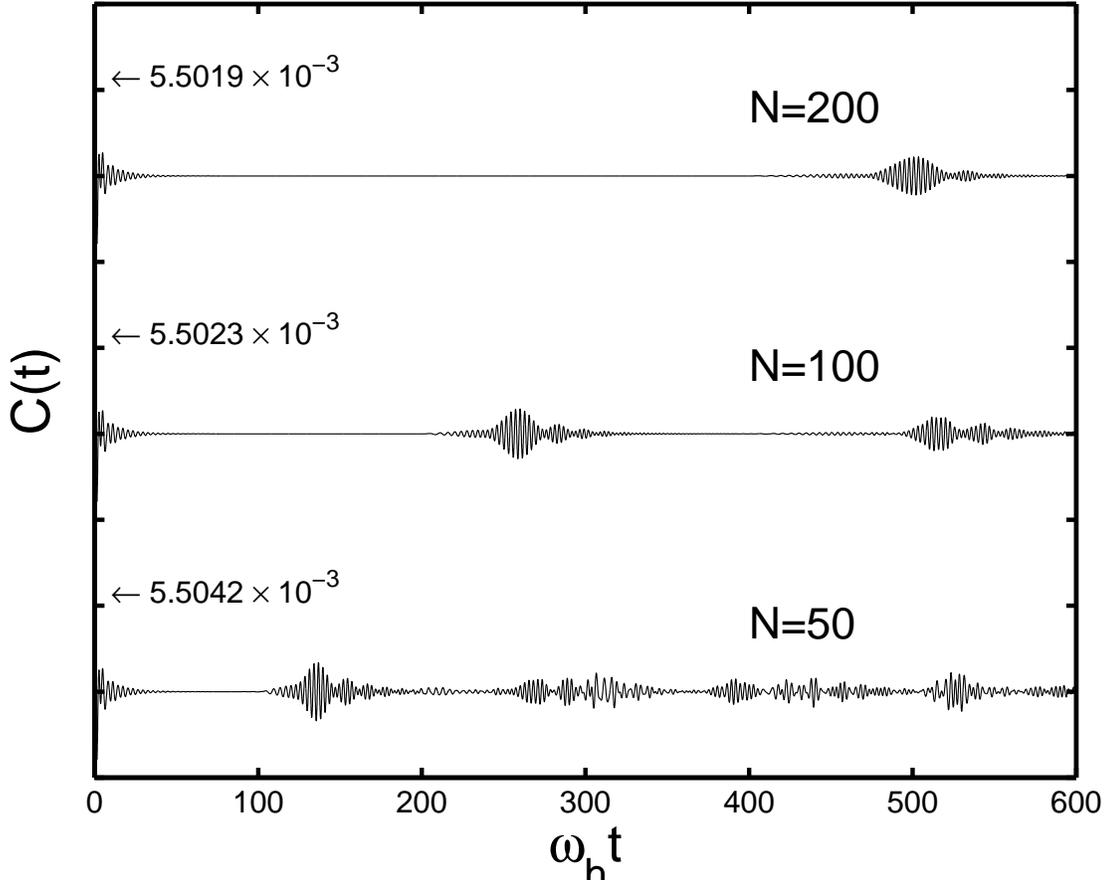}
 }
 \nopagebreak
 \caption{
  Symmetrized correlator of the energy fluctuations versus time for
  $\omega_0/\omega_h = 1$, $m_h/m_0 = 0.1$, and different numbers
  of particles $N$ in a finite string
  in units with $\hbar = 1$. The left arrows indicate the
  value of the correlator at $t = 0$.
 }
\label{FIG.2}
\end{figure}
\begin{multicols}{2}\narrowtext\noindent
The amplitudes $A_k$ and eigenfrequencies $\omega_k$ were
numerically calculated and equation (\ref{ups}) was evaluated  for
$\omega_0/\omega_h = 1$, $m_h/m_0 = 0.1$, and different values of $N$. The
results are shown in Fig. \ref{FIG.2}.

Since the string is finite, we now have a recurrence phenomenon. As shown in
Fig. \ref{FIG.2}, the correlation function initially oscillates and its
envelope decreases rapidly (as it would for the infinite chain), but it
rebuilts after a time it takes a signal to travel along the string and to
return to the oscillator after reflection at the opposite end. The built-up
time is different from the decay time of the correlation and in detail the
reconstituted correlation is clearly not the same as the initial correlation.
This behavior is a consequence of the fact that the eigenfrequencies
$\omega_k$ of our problem are only nearly exact multiples of the lowest
eigenfrequency. Even in the limit of a continuous yet finite string ($m_h \to
0$, $\omega_h \to \infty$), the equidistant levels of the system are
disturbed near the test-oscillator frequency $\omega_0$. The fact that the
eigenfrequencies  are not exact multiples leads to the more complex
phenomenon depicted in Fig.~\ref{FIG.2}. The above results suggest that a
transition to the thermodynamic limit is possible if the travel time of a
perturbation through the system is longer than the duration of an experiment.

The exact diagonalization demonstrates that the energy of a small
system coupled to a bath fluctuates as a function of time
despite the fact that the overall energy of the system is fixed. This is a
consequence of the finite coupling energy between the system and the bath.
As a result the test oscillator is not in a normal mode of the
total system but as shown by Eq. (\ref{q(t)}) in a superposition of
the true normal modes of the total system. It is for this reason that
the simple argument \cite{alein,gavish}
mentioned in the introduction fails.

The results presented in this work demonstrate that even at zero
temperatures vacuum fluctuations of the bath give rise to a ground state
that exhibits a non-trivial dynamics.
Textbook statistical mechanics \cite{becker} assumes that the coupling energy
between the test system and the bath can be neglected. In contrast, here
the coupling energy is essential. The effect discussed here can not
simply be absorbed into
renormalized parameters of the test system. We believe that these
observations are
crucial in understanding the zero-temperature properties of systems
and that they are applicable not only to the particular system investigated
here but are of a very general nature.

This work was supported by the Swiss National Science Foundation and
the Russian Foundation for Basic Research, grant 01-02-17220.

\end{multicols}
\end{document}